\documentclass[a4paper,11pt]{article}
\usepackage[T1]{fontenc}
\usepackage{jheppub}

\newcommand{\CP}{C^{\mathsf P}}
\newcommand{\wP}{w^{\mathsf P}}
\newcommand{\Res}{\operatorname*{Res}}
\newcommand{\dd}{\mathrm d}

\title{Six-point consistency and uniqueness of the Veneziano amplitude}
\author{Ilmo Sung}
\affiliation{Science and Technology Directorate, U.S. Department of Homeland Security, \\
Washington, DC, USA}
\emailAdd{ilmo.sung@hq.dhs.gov}
\abstract{We establish uniqueness of the Veneziano four-point amplitude in a meromorphic class by combining six-point consistency with a known low-order supersymmetry relation. We consider planar tree-level scattering in four dimensions with a massless maximally supersymmetric vector multiplet and an additional scalar parity condition. A difference of supersymmetry Ward identities isolates massless factorization residues and eliminates every permitted contribution regular in a common kinematic limit. The resulting functional equation holds at every derivative order and fixes all angular dependence from the forward function after subtraction of the massless poles, at fixed Yang-Mills coupling. For a jointly meromorphic amplitude with simple planar poles, a nonempty spectrum of positive mass-squared poles separated from zero by a mass gap, nonnegative coefficients in the partial-wave expansions of scalar residues, and an exact unsubtracted forward dispersion relation, only the pole positions remain free. Positivity bounds their separation. The low-order relation saturates a sharp spectral inequality, requiring an infinite sequence of equally spaced mass-squared poles. The coupling and first massive pole position therefore determine the complete four-point function and all residues, without an initial finite-spin restriction.}
\keywords{Scattering Amplitudes, Supersymmetric Gauge Theory, Effective Field Theories, Superstrings and Heterotic Strings}

\begin{document}
\maketitle
\flushbottom

\section{Introduction}\label{sec:intro}
We establish uniqueness of the Veneziano four-point amplitude from six-point consistency, an established low-order supersymmetry relation, and explicit analytic and spectral assumptions. The setting is planar tree-level scattering in four dimensions with a massless $\mathcal N=4$ vector multiplet and an additional scalar parity condition. In the stipulated meromorphic class with partial-wave positivity of the scalar residues and an exact unsubtracted forward dispersion relation, the coupling and first massive pole position determine an infinite sequence of four-point poles and all their residues. Neither the remaining pole positions nor a finite-spin restriction is supplied as input.

At six points, massless factorization produces products of four-point amplitudes, while unknown five-point data and local six-point interactions also contribute. The central question is whether the stipulated consistency conditions impose a relation on four-point data that cannot be modified by these additional contributions.

We impose unbroken $SU(4)$ R symmetry and the ordinary nonzero super Yang-Mills (SYM) cubic coupling. For the color-ordered four-gluon amplitude, define
\begin{equation}\label{eq:normalization}
\begin{aligned}
 A_4(1^-,2^-,3^+,4^+)&=\langle12\rangle^2[34]^2F(s,u),\\
 F(s,u)&=-\frac{g^2}{su}+f(s,u),\\
 M(s,u)&=-\frac{su}{g^2}F(s,u)=1-\frac{su}{g^2}f(s,u).
\end{aligned}
\end{equation}
Here $s=(p_1+p_2)^2$, $u=(p_1+p_4)^2$, and $g^2>0$ is the Yang-Mills coupling squared. Angle and square brackets denote the massless spinor contractions defined in Sec.~\ref{sec:assumptions}. The function $F$ multiplies the helicity factor, $f$ is analytic near the origin, and $M$ obeys $M(s,0)=M(0,u)=1$. The six-point conditions imply
\begin{equation}\label{eq:functional}
 M(x,y)M(x+y,z)=M(x,y+z)M(y,z)
\end{equation}
for three independent complex arguments near the origin. We derive this relation at every derivative order from the specified six-point supersymmetry Ward identities, the separately imposed scalar parity condition, and massless factorization. The permitted contributions regular in the chosen factorization limit do not contribute to the extracted constraint, regardless of their coefficients. At fixed $g$, Eq.~\eqref{eq:functional} determines the regular correction $f(s,u)$ in a neighborhood of the origin from $f(s,0)$, fixing every nonforward coefficient in its low-energy expansion.

To determine the remaining forward function, we require joint meromorphy with only simple planar poles, a nonempty spectrum of massive poles at positive mass squared with a mass gap, nonnegative coefficients in the partial-wave expansions of the residues in a physical scalar channel, and an exact unsubtracted forward dispersion relation. Combining these assumptions with the derived functional equation and the established low-order supersymmetry relation stated in Sec.~\ref{sec:assumptions} gives
\begin{equation}\label{eq:answer}
 F(s,u)=-\frac{g^2}{su}
 \frac{\Gamma(1-s/\mu_1)\Gamma(1-u/\mu_1)}
 {\Gamma(1-(s+u)/\mu_1)}.
\end{equation}
Here $\mu_1>0$ is the first massive pole position and has dimension mass squared. The distinct massive poles are necessarily $\mu_n=n\mu_1$ for every positive integer $n$. No finite-spin restriction is imposed initially. The result is the Veneziano amplitude with the kinematic factor appropriate to four-gluon scattering~\cite{Veneziano}.

Positivity bounds neighboring pole separations. The incorporated low-order equality saturates the resulting spectral bound and requires every pole of this infinite sequence.

The physical starting point is the analysis of Elvang, Herderschee, and Morales~\cite{EHM}, where six-point supersymmetry, scalar parity, and massless factorization give nonlinear four-point relations through the computed derivative orders and, with positivity, numerical evidence for uniqueness of the open-superstring amplitude. Complementary numerical approaches combine positivity with string monodromy~\cite{Carving,Flattening,StringEFT} or information about the lowest massive states~\cite{Corners}. Multiparticle factorization excludes specific deformations satisfying lower-point constraints~\cite{MultiparticleRigidity,HigherSpinConstraints}. In bi-adjoint scalar effective field theories (EFTs), six-point factorization combined with Kleiss-Kuijf and Bern-Carrasco-Johansson relations also produces nonlinear four-point constraints and emergent string monodromy~\cite{EmergentMonodromy}.

An alternative higher-point input is splitting: factorization into lower-point functions on special kinematic loci away from physical poles~\cite{HiddenZeros,AllOrderSplits}. Five-point splitting gives nonlinear four-point constraints and, with additional spectral assumptions, numerically shrinking regions around the string amplitude~\cite{SplittingIslands}. Wan and Zhou~\cite{WanZhou} derive Eq.~\eqref{eq:functional} from that input and express its solution in terms of a single auxiliary function, from which they obtain a dispersive product representation. In the present argument, the functional equation follows from the specified six-point consistency conditions rather than an assumed splitting relation. Shao and Vichi~\cite{ShaoVichi} obtain a related pole-separation condition from shifted residue zeros in meromorphic product amplitudes with infinite spin sums. Veneziano asymptotics in a high-energy regime~\cite{AsymptoticUniqueness} and exact uniqueness from minimal residue zeros and high-energy softness~\cite{StringsAlmostNothing} follow from different premises.

The conclusion applies to every four-point function admitting at least one stipulated six-point completion; it does not classify those completions. Section~\ref{sec:assumptions} defines the local amplitude data. Section~\ref{sec:sixpoint} establishes forward reconstruction, Sec.~\ref{sec:spectrum} determines the amplitude under the ultraviolet assumptions, and Sec.~\ref{sec:discussion} discusses its consequences and scope.

\section{Amplitudes and six-point conditions}\label{sec:assumptions}
\subsection{Kinematics and the four-point function}
We work at planar color-ordered tree level in complexified four-dimensional massless kinematics. The massless states entering the factorization channels form the $\mathcal N=4$ vector multiplet, with unbroken $SU(4)$ R symmetry and the ordinary nonzero SYM cubic coupling. All momenta are outgoing. The spinor conventions are
\begin{equation}\label{eq:kinematics}
 p_i=\lambda_i\widetilde\lambda_i,\qquad
 \sum_i p_i=0,\qquad s_{ij}=\langle ij\rangle[ij],
\end{equation}
with both brackets defined as two-by-two determinants. Thus
$\langle ij\rangle=\lambda_i^0\lambda_j^1-\lambda_i^1\lambda_j^0$, and the same determinant convention defines $[ij]$. Three-particle invariants are $s_{ijk}=s_{ij}+s_{ik}+s_{jk}$. At four points, $(s,u,t)=(s_{12},s_{14},s_{13})$ and $s+t+u=0$. These conventions give the negative massless prefactor in Eq.~\eqref{eq:answer}.

The symmetric four-point function has the convergent expansion
\begin{equation}\label{eq:expansion}
\begin{aligned}
 F(s,u)&=-\frac{g^2}{su}+f(s,u),\\
 f(s,u)&=\sum_{k=0}^{\infty}\sum_{q=0}^k
 a_{kq}s^{k-q}u^q,\qquad a_{kq}=a_{k,k-q},
\end{aligned}
\end{equation}
in a neighborhood of the origin. The $a_{kq}$ are the four-point Wilson coefficients in this amplitude normalization. The index $k$ is the total degree in Mandelstam variables, and $q$ is the power of $u$. These coefficients are real in real kinematic conventions. The function entering the forward dispersion relation is the regular function $f(s,0)$, rather than the singular expression $F(s,0)$.

The coupling $g$ is dimensionless, with $[s]=[u]=[\mu_n]={\rm mass}^2$, $[F]={\rm mass}^{-4}$, and $[a_{kq}]={\rm mass}^{-2k-4}$. The coefficients $a_{kq}$ are unspecified before the six-point conditions are imposed.

\subsection{Six-point supersymmetry, factorization, and scalar parity}
We select six cyclic scalar components whose supersymmetry relations involve five independent coefficient functions. A linear relation among the components can therefore eliminate those functions. Let $\phi^{AB}=-\phi^{BA}$ denote a scalar with antisymmetric $SU(4)$ indices. Consider the component
\begin{equation}\label{eq:scalarone}
 Z_1=A_6(\phi^{12},\phi^{13},\phi^{14},
              \phi^{23},\phi^{34},\phi^{24})
\end{equation}
and the five components obtained by left cyclic rotation of this state list at fixed momenta. Equivalently,
\begin{equation}\label{eq:cyclic}
\begin{aligned}
 Z_j(p_1,\ldots,p_6)&=
 Z_1(p_{\sigma_j(1)},\ldots,p_{\sigma_j(6)}),\\
 \sigma_j(i)&=1+((i-j)\bmod6).
\end{aligned}
\end{equation}
The right cyclic relabeling of momenta in this equation fixes the signs and channel assignments used throughout.

The six-scalar components lie in the next-to-maximally-helicity-violating (NMHV) sector. A basis of solutions to the supersymmetry Ward identities expresses these components in terms of five independent coefficient functions~\cite{EHM,EFK}. Thus, at generic momenta satisfying momentum conservation,
\begin{equation}\label{eq:components}
 Z_j=\sum_{q=0}^4 C_{jq}U_q,\qquad j=1,\ldots,6.
\end{equation}
Appendix~\ref{app:matrix} constructs $C$ from the scalar index assignments and Grassmann signs. We use its numerator form, absorbing the common basis denominator into the $U_q$ on a patch where that denominator is nonzero. The five coefficient functions are otherwise unrestricted. Locality and parity are conditions on the full scalar amplitudes, not on the individual $U_q$.

In addition to supersymmetry, we require each selected scalar amplitude, as a function of the external momenta, to be invariant under simultaneous exchange of angle and square brackets:
\begin{equation}\label{eq:parity}
 Z_j(\lambda,\widetilde\lambda)
 =Z_j(\widetilde\lambda,\lambda).
\end{equation}
The scalar labels are fixed, and the bracket exchange is algebraic, without complex conjugation. It excludes parity-odd Levi-Civita contractions in these scalar amplitudes and is separate from maximal supersymmetry, $SU(4)$ R symmetry, and parity acting also on the external states.

In the scalar-sector interpretation of Ref.~\cite{EHM}, Sec.~1.1, simultaneous conjugation of the three complex scalars reverses three real scalar directions. This determinant-$-1$ transformation extends $SO(6)$ to $O(6)$ in the pure scalar sector. We impose precisely Eq.~\eqref{eq:parity} on the selected components, allowing the original and bracket-exchanged Ward relations to constrain the same scalar vector. Related parity conditions on selected scalar amplitudes have also been studied in maximally supersymmetric gravitational EFTs~\cite{GravitationalParity}.

Four-point supersymmetry gives $\mathcal A_4=M(s,u)\mathcal A_{4,\mathrm{SYM}}$ for the superamplitude~\cite{EHM}, Eq.~(2.11). Each multiplier is independent of the internal state. Massless three-particle factorization therefore multiplies the SYM residue by the two factors $M$ evaluated on the corresponding four-point kinematics.

For the scalar ordering in Eq.~\eqref{eq:scalarone}, the SYM component is given by Ref.~\cite{EHM}, Eq.~(3.11):
\begin{equation}\label{eq:sym}
 Z_{1,\mathrm{SYM}}=g^4\left(
 \frac1{s_{34}}-\frac{s_{24}}{s_{34}s_{234}}
 -\frac{s_{35}}{s_{34}s_{345}}\right).
\end{equation}
For example, factorization on $s_{234}=0$ gives
\begin{equation}\label{eq:factorization234}
\begin{aligned}
 \left.s_{234}Z_1\right|_{s_{234}=0}
 ={}&-\frac{g^4s_{24}}{s_{34}}M(s_{23},s_{34})\\
 &\quad\times M(s_{16},s_{56})\Big|_{s_{234}=0}.
\end{aligned}
\end{equation}
The second pole term in Eq.~\eqref{eq:sym} gives the $s_{345}$ residue. These identities fix residues on their pole divisors; choosing the displayed expressions as pole-term representatives gives
\begin{equation}\label{eq:poles}
\begin{aligned}
 \frac{Z_1}{g^4}={}&-\frac{s_{24}}{s_{34}s_{234}}
 M(s_{23},s_{34})M(s_{16},s_{56})\\
 &-\frac{s_{35}}{s_{34}s_{345}}
 M(s_{34},s_{45})M(s_{12},s_{16})
 +\frac{\mathcal R_1}{g^4}.
\end{aligned}
\end{equation}
The other components follow by Eq.~\eqref{eq:cyclic}. The products include the SYM term and all contributions linear and quadratic in $f$. At each homogeneous order in the EFT expansion, the remainder $\mathcal R_j$ consists of local polynomial terms and the allowed terms with adjacent two-particle poles. All permitted five-point data and local six-point interactions remain arbitrary. There are no nonadjacent massless poles and no double pole in overlapping three-particle channels such as $s_{234}$ and $s_{345}$. The overlapping channels cannot occur together in an ordinary planar tree diagram.

The component relation~\eqref{eq:components}, scalar parity~\eqref{eq:parity}, and decomposition~\eqref{eq:poles} hold coefficientwise at every derivative order with the stated remainder class. Only the four-point expansion~\eqref{eq:expansion} must converge; the six-point expansion is formal. The argument applies to every $F$ admitting at least one such completion, without selecting five-point coefficients or six-point local terms.

We also impose the established low-order relation in Eq.~(1.11) of Ref.~\cite{EHM},
\begin{equation}\label{eq:loworder}
 g^2a_{20}=\frac25a_{00}^2.
\end{equation}
Here $a_{00}$ and $a_{20}$ multiply $1$ and $s^2$ in $f(s,0)=\sum_{k\ge0}a_{k0}s^k$, with the normalization of Ref.~\cite{EHM}, Eqs.~(1.2)-(1.3). In that analysis, the residues at five points are fixed by factorization into three- and four-point amplitudes (Sec.~2.3). At six points, three-particle channels factorize into two four-point amplitudes, whereas adjacent two-particle channels factorize into three- and five-point amplitudes (Secs.~3.2-3.3). Equation~\eqref{eq:loworder} is retained explicitly as an input for spectral saturation. The all-order reconstruction uses only the component, scalar-parity, and remainder conditions above.

\section{Four-point reconstruction from six-point consistency}\label{sec:sixpoint}
Pole limits of higher-point amplitude relations constrain four-point coefficients in bi-adjoint scalar EFTs~\cite{EmergentMonodromy}. Here we subtract scalar supersymmetry Ward identities and their bracket-exchanged counterparts. In a common kinematic limit, their difference vanishes linearly while the prescribed three-particle terms have simple poles. The constant Laurent coefficient then contains only factorization residues, independently of all permitted regular contributions.

\subsection{A common kinematic limit of the Ward identities}
Write $Z=(Z_1,\ldots,Z_6)^{\mathsf T}$. The five functions in Eq.~\eqref{eq:components} can be eliminated by a left null vector of $C$. Let $\CP$ denote the component matrix with angle and square brackets exchanged. On an open kinematic region where $C$ and $\CP$ have rank five, define row vectors $w$ and $\wP$ by
\begin{equation}\label{eq:nullrows}
 wC=0,\qquad \wP\CP=0,\qquad w_1=\wP_1=1.
\end{equation}
Supersymmetry and scalar parity imply
\begin{equation}\label{eq:ward}
 wZ=0,\qquad \wP Z=0.
\end{equation}
Both equations act on the same six-component scalar vector. Their difference isolates pole information where the normalized rows coincide but have different first derivatives.

Consider the conserved complex locus
\begin{equation}\label{eq:locus}
\begin{gathered}
 p_1=p_3,\qquad s_{234}=s_{345}=0,\qquad s_{123}=2y,\\
 (s_{12},s_{23},s_{34},s_{45},s_{56},s_{16})\\
 =(y,y,x,y+z,x+y,z).
\end{gathered}
\end{equation}
The six adjacent invariants and $2y$ are nonzero on the patch used below. On this locus the normalized rows take the invariant form
\begin{equation}\label{eq:rowlimit}
 w=\wP=\left(1,\frac{y}{y+z},-\frac{yz}{x(y+z)},
 -\frac{(x+y)z}{x(y+z)},0,0\right).
\end{equation}
Appendix~\ref{app:kinematics} gives a conserved spinor parametrization and verifies the common-row identities. At $(x,y,z)=(1,2,-1/3)$, both submatrices formed from rows 2 through 6 are nonsingular and both supersymmetry-basis denominators are nonzero. The normalized rows are therefore regular rational functions in a neighborhood, and both component matrices retain rank five on a nonempty open patch. The constraint below follows from the difference between the first derivatives of the two normalized null vectors along a transverse deformation, rather than from their common value on the locus.

\subsection{Isolation of the massless pole contributions}
Choose a rational one-parameter deformation of the spinors that preserves momentum conservation and passes through the regular part of the locus at $\tau=0$, such that
\begin{equation}\label{eq:deformation}
\begin{aligned}
 s_{234}&=\alpha\tau+O(\tau^2),\\
 s_{345}&=\beta\tau+O(\tau^2),\\
 w-\wP&=\tau\nu+O(\tau^2),\qquad \alpha\beta\ne0.
\end{aligned}
\end{equation}
Appendix~\ref{app:kinematics} specifies this deformation by varying the spinors and solving momentum conservation exactly. The adjacent two-particle invariants remain nonzero at $\tau=0$. Every allowed local term and every allowed term with an adjacent two-particle pole in $\mathcal R_j$ is consequently regular in $\tau$. In particular, the remainder has no pole at the nonadjacent invariant $s_{13}=0$.

The absence of an overlapping-channel double pole gives
\begin{equation}\label{eq:laurentexpansion}
 Z=\tau^{-1}Z_{-1}+Z_0+O(\tau).
\end{equation}
Taking the constant Laurent coefficient of the difference of Eq.~\eqref{eq:ward} yields
\begin{equation}\label{eq:laurent}
 0=[\tau^0]\bigl((w-\wP)Z\bigr)=\nu Z_{-1}.
\end{equation}
This coefficient contains only the specified three-particle residues. Every permitted local term and every contribution with an adjacent two-particle pole is multiplied by a row difference of order $\tau$ and therefore does not contribute to this coefficient. The cancellation holds for arbitrary allowed remainders at every derivative order.

On the locus, introduce
\begin{equation}\label{eq:products}
\begin{aligned}
 \mathcal A&=M(x,y)M(x+y,z),\\
 \mathcal B&=M(x,y+z)M(y,z).
\end{aligned}
\end{equation}
Cyclic relabeling of the two terms in Eq.~\eqref{eq:poles} gives Table~\ref{tab:poles}. Eight of the twelve terms contribute to $Z_{-1}$. The remaining four have three-particle denominator $2y$, which is nonzero. Symmetry of $M$ identifies the surviving argument pairs with precisely the two products in Eq.~\eqref{eq:products}.

\begin{table}[tb]
\caption{Coefficients in $(Z_{-1})_j/g^4=c_{j\mathcal A}\mathcal A+c_{j\mathcal B}\mathcal B$ on the locus of Eq.~\eqref{eq:locus}. The quantities $\alpha$ and $\beta$ are the channel derivatives in Eq.~\eqref{eq:deformation}. All twelve cyclic source terms are included.}
\label{tab:poles}
\renewcommand{\arraystretch}{1.65}
\centering
\begin{tabular}{ccc}
\hline
Component $j$ & $c_{j\mathcal A}$ & $c_{j\mathcal B}$\\
\hline
1 & $\frac{x + y}{\alpha x}$ & $\frac{x + y + z}{\beta x}$ \\
2 & $\frac{x + y}{\alpha y}$ & $0$ \\
3 & $0$ & $\frac{y + z}{\beta y}$ \\
4 & $\frac{x + y + z}{\alpha z}$ & $\frac{y + z}{\beta z}$ \\
5 & $\frac{x + y + z}{\alpha \left(x + y\right)}$ & $0$ \\
6 & $0$ & $\frac{x + y + z}{\beta \left(y + z\right)}$ \\
\hline
\end{tabular}

\end{table}

Consequently, Eq.~\eqref{eq:laurent} becomes
\begin{equation}\label{eq:projected}
\begin{gathered}
 0=g^4\bigl(K_{\mathcal A}\mathcal A+
 K_{\mathcal B}\mathcal B\bigr),\\
 K_{\mathcal A}=\sum_j\nu_jc_{j\mathcal A},\qquad
 K_{\mathcal B}=\sum_j\nu_jc_{j\mathcal B}.
\end{gathered}
\end{equation}
The coefficients are rational functions of the base kinematics and deformation direction. They are independent of the unknown four-point amplitude.

The SYM component in Eq.~\eqref{eq:sym} obeys both Ward identities and has $M=1$. Extracting the same Laurent coefficient from this established solution gives the kinematic identity
\begin{equation}\label{eq:relative}
 K_{\mathcal A}+K_{\mathcal B}=0.
\end{equation}
This determines the relative coefficient without using any higher-derivative four-point data.

For the explicit deformation in Appendix~\ref{app:kinematics}, evaluated at $(x,y,z)=(1,2,-1/3)$, exact rational linear algebra gives
\begin{equation}\label{eq:nonzero}
 K_{\mathcal A}=-\frac45,\qquad
 K_{\mathcal B}=\frac45.
\end{equation}
Hence $K_{\mathcal A}$ is a nonzero rational function. On its nonvanishing open set, Eq.~\eqref{eq:projected} and $g^2>0$ imply $\mathcal A=\mathcal B$. Table~\ref{tab:poles} establishes the generic two-product form, the SYM solution fixes the relative coefficient, and the rational evaluation establishes nonvanishing.

\subsection{Extension to every order in the derivative expansion}\label{subsec:allorders}
It remains to pass from the selected kinematics to three independent arguments. Let $\eta,q$ be the two free parameters of the momentum-conserving spinor parametrization in Appendix~\ref{app:kinematics}, and let $\rho$ restore its overall invariant scale. The resulting map is
\begin{equation}\label{eq:dominance}
\begin{aligned}
 (x,y,z)&=\left(\rho,\rho\eta,
 -\frac{\rho\eta(1+\eta)}{q^2+\eta}\right),\\
 \det\frac{\partial(x,y,z)}{\partial(\rho,\eta,q)}
 &=\frac{2\rho^2\eta(1+\eta)q}{(q^2+\eta)^2}.
\end{aligned}
\end{equation}
The Jacobian is a nonzero rational function. The image of the regular parameter domain therefore contains an open set in the three complex variables $x,y,z$.

At a fixed homogeneous order, the difference between the two sides of Eq.~\eqref{eq:functional} is a polynomial in $x,y,z$. Its pullback vanishes on the nonempty open set where the parametrization and normalized rows are regular, the two channel derivatives are nonzero, and $K_{\mathcal A}\ne0$. The image contains an open set, so this polynomial vanishes identically.

The Ward matrices, the pole assignments, and the elimination of regular terms are independent of the EFT order. The same reasoning therefore applies at every homogeneous degree and proves Eq.~\eqref{eq:functional} as a formal power-series identity. Convergence of the four-point expansion promotes it to an analytic identity after the neighborhood is reduced so that all argument pairs lie within the convergence domain. Only the four-point convergence is used in this last step; the six-point conditions remain coefficientwise.

\subsection{Reconstruction from the regular forward function}\label{subsec:forward}
The normalized analytic solutions of Eq.~\eqref{eq:functional} can be obtained directly~\cite{WanZhou}. Since $M(0,0)=1$, there is a neighborhood of the origin in which $M$ is nonzero. Let $L=\log M$, choosing the branch with $L(0,0)=0$, and let $\partial_2$ denote differentiation with respect to the second argument. Define $H'(s)=-\partial_2L(s,0)$ and $H(0)=0$. Differentiating the additive equation in its third argument at zero gives
\begin{equation}\label{eq:logdifferentiated}
 \partial_2L(s,u)=H'(u)-H'(s+u).
\end{equation}
Integration in $u$, using $L(s,0)=0$, yields
\begin{equation}\label{eq:logsolution}
\begin{gathered}
 M(s,u)=\exp\!\left[H(s)+H(u)-H(s+u)\right],\\
 H(0)=H'(0)=0.
\end{gathered}
\end{equation}
The condition $H'(0)=0$ follows from $L(0,u)=0$. Adding a linear term to $H$ leaves the amplitude unchanged; the displayed normalization fixes this freedom.

From $M=1-su f/g^2$ one obtains
\begin{equation}\label{eq:forwardgenerator}
 H'(s)=\frac{s f(s,0)}{g^2},\qquad
 H(s)=\sum_{k=0}^{\infty}\frac{a_{k0}s^{k+2}}{(k+2)g^2}.
\end{equation}
At fixed $g$, Eqs.~\eqref{eq:logsolution} and~\eqref{eq:forwardgenerator} determine every nonforward low-energy coefficient from the regular forward coefficients. In particular, two four-point functions with the same coupling, the same $f(s,0)$, and the stipulated six-point completions have identical analytic germs. This is a necessary condition, not a construction of a completion for arbitrary $H$.

At quadratic order in $s,u$, Eqs.~\eqref{eq:logsolution} and~\eqref{eq:forwardgenerator} give the nonforward Wilson coefficient
\begin{equation}\label{eq:angularloworder}
 a_{21}=\frac32a_{20}-\frac{a_{00}^2}{2g^2}.
\end{equation}
The same algebraic coefficient relation follows from five-point splitting in Ref.~\cite{SplittingIslands}, Eq.~(1.8); here it follows from the six-point conditions. The functional equation fixes this nonforward coefficient while leaving $a_{20}$ free. Combining Eq.~\eqref{eq:angularloworder} with Eq.~\eqref{eq:loworder} gives $g^2a_{21}=a_{00}^2/10$. No massive spectrum, residue expansion, or dispersion relation has entered the reconstruction. The ultraviolet conditions will now restrict the forward function.

\section{Meromorphy, positivity, and the Veneziano amplitude}\label{sec:spectrum}
It remains to determine the regular forward function. Meromorphy specifies its singularities, partial-wave positivity constrains the residues, and forward dispersion relates those residues to the function.

\subsection{Analytic and spectral assumptions}

We require the four-point function to extend jointly meromorphically to $\mathbb C^2$. Its only possible polar hypersurfaces are
\begin{equation}\label{eq:polarset}
 s=0,\qquad u=0,\qquad s=\mu_n,\qquad u=\mu_n.
\end{equation}
The distinct genuine massive pole positions form a nonempty finite or countable strictly increasing positive sequence, with no finite accumulation and smallest element $\mu_1$. Simplicity is imposed separately in each variable; mixed poles at intersections of massive $s$- and $u$-channel divisors are permitted initially, and the residue functions are not assumed polynomial. There are no cuts or additional polar divisors. The massless residues are exactly those of SYM, and $f$ is regular on the massless axes away from the massive poles. These assumptions concern the analytic four-point function, not merely its formal low-energy expansion.

For a genuine massive pole, define the residue with orientation
\begin{equation}\label{eq:resdef}
 R_n(u)=\lim_{s\to\mu_n}(\mu_n-s)F(s,u).
\end{equation}
In the channel $A_4(zz\bar z\bar z)=s^2F(s,u)$, where $z$ and $\bar z$ denote a conjugate pair of complex scalars~\cite{EHM}, the physical residue is $\mu_n^2R_n(u)$. The prefactor is positive and independent of the angle; $[R_n]={\rm mass}^{-2}$. On the physical interval $-\mu_n\le u\le0$, require
\begin{equation}\label{eq:positivity}
 R_n(u)=\sum_{\ell=0}^{\infty}b_{n\ell}
 P_\ell\!\left(1+\frac{2u}{\mu_n}\right),
 \qquad b_{n\ell}\ge0,
\end{equation}
where $P_\ell$ is the degree-$\ell$ Legendre polynomial, normalized by $P_\ell(1)=1$, and at least one coefficient is nonzero at each pole. The expansion represents the actual analytic residue, with real analyticity in the stated conventions. We allow infinitely many contributing spins. The continuation needed outside the physical interval is established below.

Finally, set $r_n=R_n(0)>0$ and impose the exact forward identity
\begin{equation}\label{eq:dispersion}
 f(s,0)=\sum_n\frac{r_n}{\mu_n-s},
 \qquad \sum_n\frac{r_n}{\mu_n}<\infty.
\end{equation}
The series converges absolutely and uniformly on compact subsets of the complex $s$-plane that do not contain poles. There is no independent polynomial, entire-function, subtraction, or large-contour contribution. Equation~\eqref{eq:dispersion} is a boundary condition on the regular forward function, in addition to meromorphy and partial-wave positivity.

For positively oriented circles $|z|=R_j$ avoiding the poles and enclosing a fixed regular point $s$, Cauchy's theorem gives
\begin{equation}\label{eq:forwardcontour}
\begin{aligned}
 f(s,0)&=\sum_{\mu_n<R_j}\frac{r_n}{\mu_n-s}+I_j(s),\\
 I_j(s)&=\frac{1}{2\pi i}\oint_{|z|=R_j}
             \frac{f(z,0)}{z-s}\,\dd z.
\end{aligned}
\end{equation}
A sufficient condition for $I_j(s)\to0$ is $\sup_{|z|=R_j}|f(z,0)|\to0$ along a sequence $R_j\to\infty$.\footnote{For fixed $s$ with $|s|<R_j$, the contour term obeys $|I_j(s)|\leq R_j(R_j-|s|)^{-1}\sup_{|z|=R_j}|f(z,0)|$. The summability in Eq.~\eqref{eq:dispersion} then gives the pole sum in the limit.} This is sufficient, not an additional premise of the classification. A bound at fixed negative transfer does not by itself establish this zero-transfer condition without control of the forward limit.

\subsection{Simple poles and the product representation}
The dispersive product and the unit exponents of simple poles appear in Ref.~\cite{WanZhou}, Eqs.~(39)-(41). We derive the representation from Eq.~\eqref{eq:dispersion}, fixing the forward residues by a meromorphic logarithmic derivative before constructing the global auxiliary function.

Set $\chi(s)=s f(s,0)/g^2$, a globally meromorphic function. Equations~\eqref{eq:logsolution} and~\eqref{eq:forwardgenerator} imply locally
\begin{equation}\label{eq:logderivative}
 \frac{\partial_sM(s,u)}{M(s,u)}
 =\chi(s)-\chi(s+u).
\end{equation}
Both sides have meromorphic continuations supplied by $F$ and Eq.~\eqref{eq:dispersion}. The meromorphic identity theorem extends this equality to their full domain. This step does not require a globally single-valued $H$.

At a genuine massive pole $s=\mu_n$, choose a sufficiently small generic nonzero $u$ such that $R_n(u)\ne0$ and $\mu_n+u$ is not another pole. Such a choice exists because $R_n$ is analytic near zero and $R_n(0)=r_n>0$. The left side of Eq.~\eqref{eq:logderivative} is the logarithmic derivative of a function with a simple pole and therefore has residue $-1$. The shifted term is regular there. The forward identity gives
\begin{equation}\label{eq:weightresidue}
 -1=\Res_{s=\mu_n}\chi(s)
 =-\frac{\mu_n r_n}{g^2}.
\end{equation}
It follows that
\begin{equation}\label{eq:moments}
 r_n=\frac{g^2}{\mu_n},\qquad
 \frac{a_{k0}}{g^2}=\sum_n\mu_n^{-k-2}.
\end{equation}
The interpretation of forward coefficients as moments of a positive spectral measure is central to dispersive positivity~\cite{PositiveMoments}. Here the simple-pole condition additionally fixes every weight. For the second identity in Eq.~\eqref{eq:moments}, expand each term in Eq.~\eqref{eq:dispersion} geometrically on a closed disk $|s|\le r<\mu_1$. The series is dominated by a fixed multiple of $\sum_n r_n/\mu_n$, which justifies exchanging the two sums.

In particular, $\sum_n\mu_n^{-2}=a_{00}/g^2<\infty$. We can now define the single-valued meromorphic function
\begin{equation}\label{eq:hproduct}
 h(s)=\prod_n\frac{e^{-s/\mu_n}}{1-s/\mu_n}.
\end{equation}
For $s$ in a compact set and all sufficiently large $n$,
$-s/\mu_n-\log(1-s/\mu_n)=O(\mu_n^{-2})$ uniformly. The product consequently converges normally away from its poles. It has no zeros, has a simple pole at every $\mu_n$, and satisfies $h(0)=1$ and $h'(0)=0$. Its logarithmic derivative is
\begin{equation}\label{eq:hlogderivative}
 \frac{h'(s)}{h(s)}
 =\sum_n\left(\frac1{\mu_n-s}-\frac1{\mu_n}\right)
 =\chi(s).
\end{equation}
The constructed function agrees with $e^{H(s)}$ near the origin: their logarithmic derivatives and normalizations coincide. Therefore
\begin{equation}\label{eq:canonicalproduct}
 M(s,u)=\frac{h(s)h(u)}{h(s+u)},
\end{equation}
first near the origin and then meromorphically. The four-point function is now determined by its pole positions. The remaining partial-wave positivity and completion conditions restrict which sequences can occur.

The role of the exact forward identity can also be seen within the reconstructed functional class. Replacing it by
\begin{equation}\label{eq:entireforward}
 f_E(s,0)=\sum_n\frac{r_n}{\mu_n-s}+E(s),
\end{equation}
with $E$ entire would change the normalized amplitude to
\begin{equation}\label{eq:entiregenerator}
\begin{aligned}
 P(s)&=\frac1{g^2}\int_0^s zE(z)\,\dd z,\\
 M_E(s,u)&=M(s,u)\exp\!\left[P(s)+P(u)-P(s+u)\right].
\end{aligned}
\end{equation}
Equation~\eqref{eq:dispersion} excludes this contribution. An arbitrary $E$ need not preserve partial-wave positivity of the residues or admit the required six-point completion. This freedom changes the regular forward function; deformations that leave it unchanged are addressed in Sec.~\ref{sec:discussion}.

\subsection{Positivity and separation of the poles}
To constrain pole separations, we first establish strict positivity of each residue at positive momentum transfer below the first massive pole in the transfer channel.

The meromorphic polar set makes $R_n(u)$ holomorphic near the compact physical interval $[-\mu_n,0]$ and throughout $|u|<\mu_1$. At the physical endpoint, Eq.~\eqref{eq:positivity} gives
$\sum_\ell b_{n\ell}=r_n<\infty$. Since $|P_\ell(z)|\le1$ on $[-1,1]$, the physical series converges uniformly and its coefficients are the ordinary orthogonal Legendre coefficients.

Holomorphy in an ellipse surrounding the physical interval implies exponential decay of those coefficients and normal convergence on a smaller ellipse. Appendix~\ref{app:analytic} gives the required bounds. In particular, derivatives can be taken term by term near $u=0$. The exact endpoint expansion is
\begin{equation}\label{eq:legendre}
 P_\ell\!\left(1+\frac{2u}{\mu_n}\right)
 =\sum_{k=0}^{\ell}
 \frac{(\ell+k)!}{(\ell-k)!(k!)^2}
 \left(\frac{u}{\mu_n}\right)^k.
\end{equation}
Every Taylor coefficient of $R_n$ at zero is consequently nonnegative. Its constant term is strictly positive, and its Taylor series converges throughout $|u|<\mu_1$. Hence
\begin{equation}\label{eq:positivegap}
 R_n(u)>0,\qquad 0<u<\mu_1.
\end{equation}
This conclusion applies to infinite spin sums and requires no continuation across a massive transfer pole.

Let
\begin{equation}\label{eq:kappa}
 \kappa_n=\lim_{s\to\mu_n}(\mu_n-s)h(s)\ne0.
\end{equation}
The product representation gives
\begin{equation}\label{eq:residueproduct}
 R_n(u)=-\frac{g^2\kappa_n}{\mu_n u}
 \frac{h(u)}{h(\mu_n+u)}.
\end{equation}
If $0<\mu_{n+1}-\mu_n<\mu_1$, then at
$u=\mu_{n+1}-\mu_n$ the numerator is finite and nonzero, while the reciprocal of the denominator has a zero. Equation~\eqref{eq:residueproduct} would give $R_n(u)=0$ inside the interval in Eq.~\eqref{eq:positivegap}. Thus
\begin{equation}\label{eq:spacing}
 \mu_{n+1}-\mu_n\ge\mu_1,\qquad
 \mu_n\ge n\mu_1.
\end{equation}
The same pole-separation argument, based on zeros of the residues, appears in meromorphic splitting amplitudes~\cite{ShaoVichi}. Here it applies to the specified physical scalar residue, including initially infinite spin sums. Equation~\eqref{eq:spacing} is necessary for partial-wave positivity, but does not suffice to establish it.

At the boundary $\mu_{n+1}-\mu_n=\mu_1$, the numerator in Eq.~\eqref{eq:residueproduct} also has a pole. The possible cancellation is allowed. Only separations strictly below $\mu_1$ have been excluded.

\subsection{Saturation of the spectral bound}
The low-order relation now constrains only the pole positions. Equation~\eqref{eq:moments} expresses its dimensionless ratio as
\begin{equation}\label{eq:ratio}
\frac{\sum_n\mu_n^{-4}}{(\sum_n\mu_n^{-2})^2}
 =\frac{g^2a_{20}}{a_{00}^2}.
\end{equation}
To bound this ratio and determine its equality condition, introduce
\begin{equation}\label{eq:sequence}
\begin{gathered}
 \xi_n=\left(\frac{\mu_1}{\mu_n}\right)^2,\qquad
 \xi_1=1,\qquad 0\le\xi_n\le\frac1{n^2},\\
 S=\sum_n\xi_n,\qquad T=\sum_n\xi_n^2.
\end{gathered}
\end{equation}
For a finite spectrum the sequence is padded by zeros. It suffices to minimize $T/S^2$ over the larger coordinate domain $\xi_1=1$, $0\le\xi_n\le n^{-2}$, which contains every spectrum satisfying Eq.~\eqref{eq:spacing}. This enlargement permits independent coordinate variations. Both sums converge absolutely. On this entire domain, $S\le\zeta(2)<2$ and $T\ge1$, where $\zeta$ is the Riemann zeta function.

Holding the other coordinates fixed and varying any $\xi_n$ with $n\ge2$ gives
\begin{equation}\label{eq:monotonicity}
 \frac{\partial}{\partial\xi_n}\frac{T}{S^2}
 =\frac{2(\xi_nS-T)}{S^3}<0.
\end{equation}
Indeed, $\xi_nS\le\zeta(2)/4<1\le T$. This strict sign holds throughout every allowed coordinate interval, including after other coordinates have been increased.

Replace successively $\xi_2,\ldots,\xi_N$ by their upper bounds. The ratio $T/S^2$ cannot increase. The summable bounds $n^{-2}$ and $n^{-4}$ allow dominated convergence as $N\to\infty$, giving
\begin{equation}\label{eq:bound}
 \frac{T}{S^2}\ge
 \frac{\zeta(4)}{\zeta(2)^2}=\frac25.
\end{equation}
The equality condition follows from the strict derivative. If one coordinate obeys $\xi_j<j^{-2}$, increasing that coordinate first produces a strict decrease. Subsequent replacements cannot increase the ratio. The original ratio is therefore strictly greater than $2/5$. Equality in Eq.~\eqref{eq:bound} holds if and only if $\xi_n=n^{-2}$ for every $n$. In particular, no finite spectrum can attain it.

By Eq.~\eqref{eq:ratio}, the ratio $T/S^2$ equals $g^2a_{20}/a_{00}^2$.
Thus the derived functional equation, meromorphy, partial-wave positivity, and the forward dispersion relation imply the sharp inequality
\begin{equation}\label{eq:wilsonbound}
 g^2a_{20}\ge\frac25a_{00}^2.
\end{equation}
The low-order supersymmetry relation supplies equality and requires
\begin{equation}\label{eq:lattice}
 \mu_n=n\mu_1,\qquad n=1,2,\ldots.
\end{equation}
Thus the low-order equality fixes every pole of an infinite spectrum after the preceding consistency conditions have restricted the spectral problem. Appendix~\ref{app:alternatives} gives four-point functions illustrating the separate roles of this equality and exact forward dispersion.

\subsection{The amplitude and its residues}
With the spectrum in Eq.~\eqref{eq:lattice}, Euler's product gives
\begin{equation}\label{eq:euler}
 h(s)=e^{-\gamma s/\mu_1}\Gamma(1-s/\mu_1),
\end{equation}
where $\gamma$ is Euler's constant. The linear exponential cancels in Eq.~\eqref{eq:canonicalproduct}, yielding Eq.~\eqref{eq:answer}. The coupling and the first massive pole position are the only undetermined quantities. In the usual string normalization, $\alpha'=1/\mu_1$.

The Veneziano forward function follows from $f(s,0)=g^2h'(s)/(s h(s))$ and Eq.~\eqref{eq:euler}:
\begin{equation}\label{eq:venezianoforward}
\begin{aligned}
 f_{\mathrm V}(s,0)
 &=-\frac{g^2}{\mu_1s}
 \left[\psi\!\left(1-\frac{s}{\mu_1}\right)+\gamma\right]\\
 &=\sum_{n=1}^{\infty}
 \frac{g^2}{n\mu_1(n\mu_1-s)}.
\end{aligned}
\end{equation}
Here $\psi(z)=\Gamma'(z)/\Gamma(z)$ is the digamma function. The apparent singularity at $s=0$ is removable, with $f_{\mathrm V}(0,0)=g^2\zeta(2)/\mu_1^2$. The normally convergent sum realizes Eq.~\eqref{eq:dispersion} with $r_n=g^2/(n\mu_1)$. Its Taylor expansion gives
\begin{equation}\label{eq:forwardcoefficients}
 a_{k0}=\frac{g^2\zeta(k+2)}{\mu_1^{k+2}}.
\end{equation}
Substitution into Eqs.~\eqref{eq:logsolution} and~\eqref{eq:forwardgenerator} fixes all remaining coefficients.

Taking the $(n\mu_1-s)$ residue and applying $\Gamma(z+1)=z\Gamma(z)$ gives
\begin{equation}\label{eq:stringresidues}
 R_n(u)=\frac{g^2}{\mu_1 n!}
 \prod_{j=1}^{n-1}\left(j+\frac{u}{\mu_1}\right),\qquad
 R_n(0)=\frac{g^2}{n\mu_1}.
\end{equation}
The residue signs follow from the same recurrence.\footnote{The $(n\mu_1-s)$ residue of $\Gamma(1-s/\mu_1)$ is $\mu_1(-1)^{n-1}/(n-1)!$. With $b=u/\mu_1$, one has $\Gamma(1-b)/\Gamma(1-n-b)=(-1)^n b\prod_{j=1}^{n-1}(b+j)$. Combining these factors with $-g^2/(su)$ gives Eq.~\eqref{eq:stringresidues}.} The empty product at $n=1$ equals one. The first scalar residue is constant, and the residue at level $n$ is a polynomial of degree $n-1$. Finite spin in this scalar channel is therefore a consequence, not an initial restriction.

The ratio of gamma functions is symmetric under $s\leftrightarrow u$ and equals one on either massless axis, reproducing the SYM residues. At fixed regular $s,u$, its limit as $\mu_1\to\infty$ is one, so $F\to-g^2/(su)$. The limiting pure-SYM function is a boundary of the problem; the admissible class itself has a nonempty massive spectrum.

The standard open-superstring disk amplitude supplies a realization with the required massless supersymmetry, scalar parity, and factorization properties~\cite{EHM}. The ratio of gamma functions has the stipulated pole set and residues, and Eq.~\eqref{eq:venezianoforward} gives its forward dispersion relation. All-level partial-wave positivity in four dimensions follows directly from Ref.~\cite{TreeUnitarity}, Sec.~4.1; the proof covering dimensions up to ten is given in Ref.~\cite{Mansfield}. The classified four-point set is therefore nonempty.

\section{Discussion and conclusions}\label{sec:discussion}
The two stages of the argument constrain different parts of the four-point amplitude. Six-point consistency determines angular dependence from the regular forward function at fixed coupling, independently of the permitted regular higher-point interactions. The ultraviolet conditions fix the forward residue weights and constrain pole separation. The established low-order supersymmetry equality then requires every pole of the infinite spectrum. The six-point conditions are imposed at every derivative order; the determination is not based on a finite Taylor truncation.

The reconstruction excludes nontrivial changes of angular dependence even when the complete regular forward function and all pole residues are unchanged. Let $F_{\mathrm V}$ denote Eq.~\eqref{eq:answer}, set $t=-s-u$, and consider
\begin{equation}\label{eq:sinedeformation}
\begin{aligned}
 F_\epsilon&=F_{\mathrm V}(1+\epsilon D_N),\\
 D_N&=\left[
 \sin\frac{\pi s}{\mu_1}\,
 \sin\frac{\pi u}{\mu_1}\,
 \sin\frac{\pi t}{\mu_1}
 \right]^{2N},
\end{aligned}
\end{equation}
where $N\ge1$ is an integer and $\epsilon\ne0$ is real. The sine zeros cancel every pole of $F_\epsilon-F_{\mathrm V}$, so the correction is jointly entire. All massive and massless residues are unchanged, and the regular forward functions satisfy $f_\epsilon(s,0)=f_{\mathrm V}(s,0)$. The first changed term in $f$ has total degree $6N-2$. Any prescribed finite derivative truncation can therefore agree with $F_{\mathrm V}$ for a sufficiently large $N$. For generic nonzero transfer, the sine multiplier grows exponentially in some complex directions; the present classification imposes no corresponding nonforward high-energy bound.

If $F_\epsilon$ admitted the stipulated six-point completion, Eq.~\eqref{eq:forwardgenerator} would give the same $H'(s)$ as for $F_{\mathrm V}$. The normalization and Eq.~\eqref{eq:logsolution} would then require $M_\epsilon=M_{\mathrm V}$ near the origin, contradicting the nonzero deformation. Appendix~\ref{app:alternatives} gives the explicit nonzero functional-equation coefficient for every $N$.

The argument is a four-point classification in the stated planar tree-level setting. Scalar parity, the allowed six-point remainder singularities, joint meromorphy, partial-wave positivity of the scalar residues, and the exact unsubtracted forward dispersion relation are substantive conditions. The result does not classify higher-point contact terms or massive-state degeneracies, and does not extend to loops or additional massless sectors. Within this class, once the Yang-Mills coupling and first massive pole position are specified, every four-point coefficient, pole position, and residue is fixed; a nontrivial change is incompatible with the stipulated six-point completion and ultraviolet conditions.

\acknowledgments
The author thanks Ryumi S. for inspiration, Yu-tin Huang for providing helpful suggestions, and Damian G. and Chris M. for useful discussions. The views expressed are those of the author and do not necessarily represent the U.S. Department of Homeland Security or the United States Government.

\appendix
\section{Six-point scalar component matrix}\label{app:matrix}
We construct the component matrix used in Sec.~\ref{sec:sixpoint}. Leg labels run from 1 to 6, and $SU(4)$ indices from 1 to 4. Let
\begin{equation}\label{eq:statelist}
 \mathcal S=((1,2),(1,3),(1,4),(2,3),(3,4),(2,4))
\end{equation}
and assign to leg $i$ of component $j$ the pair
$\mathcal S_{1+((i+j-2)\bmod6)}$. For each R index $A=1,\ldots,4$, let $T_A^{(j)}$ be the increasing triple of occupied legs. Every R index occurs on exactly three legs in this scalar component.

For $i=1,2$ define
\begin{equation}\label{eq:mdef}
 m_i(\ell)=[34]\delta_{\ell i}
 +[4i]\delta_{\ell3}+[i3]\delta_{\ell4}.
\end{equation}
For an increasing triple $T=(a,b,c)$, set
\begin{equation}\label{eq:ddef}
 d_i(T)=\langle ab\rangle m_i(c)
 -\langle ac\rangle m_i(b)+\langle bc\rangle m_i(a).
\end{equation}
This is the scalar coefficient obtained from the corresponding R-index factor in the NMHV supersymmetry basis~\cite{EHM,EFK}.

The Grassmann sign is fixed as follows. Order the occupied pairs $(\ell,A)$ first by increasing $A$ and, for fixed $A$, by increasing $\ell$. Let $\varepsilon_j$ be the sign of the permutation that instead orders these pairs first by increasing $\ell$ and, for fixed $\ell$, by increasing $A$. With auxiliary commuting variables $X,Y$, define
\begin{equation}\label{eq:binary}
 \mathcal P_j(X,Y)=\varepsilon_j
 \prod_{A=1}^4
 \left[d_1(T_A^{(j)})X+d_2(T_A^{(j)})Y\right].
\end{equation}
The component matrix is
\begin{equation}\label{eq:Cdef}
 C_{jq}=[X^qY^{4-q}]\mathcal P_j(X,Y),
 \qquad q=0,\ldots,4.
\end{equation}
The rows are ordered by $Z_j$ and the columns by ascending $q$. The common denominator of the source basis is $[34]^4\langle56\rangle^4$. Absorbing it into the coefficient functions leaves the column-span condition unchanged on its nonzero patch. An invertible constant change of the five columns also leaves the left-null equations unchanged. The exchanged matrix $\CP$ is obtained from exactly the same index and sign rules by exchanging angle and square brackets.

For example, the triples for $Z_1$ are $(1,2,3)$, $(1,4,6)$, $(2,4,5)$, and $(3,5,6)$. Their product in Eq.~\eqref{eq:binary}, with its specified permutation sign, determines the first row. Applying the cyclic rule to the state list generates all remaining rows. This construction avoids independent sign choices for the cyclic components.

\section{Conserved kinematics and evaluation of the Ward identities}\label{app:kinematics}
\subsection{Spinor parametrization and rank}
At unit invariant scale, write
\begin{equation}\label{eq:chartdefinitions}
 x=1,\qquad y=\eta,\qquad
 D=q^2+\eta,\qquad z=-\frac{\eta(1+\eta)}D.
\end{equation}
The spinors in Table~\ref{tab:spinors} satisfy momentum conservation and realize Eq.~\eqref{eq:locus}. The domain excludes zeros of the displayed denominators and of the adjacent invariants used in Sec.~\ref{sec:sixpoint}.

\begin{table}[tb]
\caption{Spinors satisfying momentum conservation for the unit-scale parametrization in Eq.~\eqref{eq:chartdefinitions}, with $D=q^2+\eta$. The two spinors of each leg define a massless outgoing momentum.}
\label{tab:spinors}
\renewcommand{\arraystretch}{1.55}
\centering
\begin{tabular}{ccc}
\hline
Leg & $\lambda_i$ & $\widetilde\lambda_i$\\
\hline
1 & $(1,0)$ & $(0,-\eta)$ \\
2 & $(0,1)$ & $(1,0)$ \\
3 & $(1,0)$ & $(0,-\eta)$ \\
4 & $(1,1)$ & $(1/\eta,1+\eta)$ \\
5 & $(q-\eta,q(1+\eta))$ & $\left(-\frac{q}{\eta D},-\frac{\eta+q}{D}\right)$ \\
6 & $(1+q,1+\eta)$ & $\left(-\frac1D,\frac{\eta(q-1)}D\right)$ \\
\hline
\end{tabular}

\end{table}

In addition to the six adjacent pairs, direct substitution gives
\begin{equation}\label{eq:nonadjacent}
\begin{aligned}
 s_{13}&=0,&s_{14}&=x,&s_{15}=s_{35}&=-x-y-z,\\
 s_{24}&=-x-y,&s_{25}&=x+z,&s_{26}&=-y-z,\\
 s_{36}&=z,&s_{46}&=-x-z.&&
\end{aligned}
\end{equation}
Together these determine all fifteen pair invariants, including $s_{234}=s_{345}=0$ and $s_{123}=2y$. Multiplying all dotted spinors by $\eta D$ clears their denominators. Each pair invariant is then multiplied by $(\eta D)^2$, and the four components of momentum conservation vanish in $\mathbb Q[\eta,q]$.

The component construction gives the common null-row identities
\begin{equation}\label{eq:commonrow}
\begin{gathered}
 v_0C=v_0\CP=0,\\
 v_0=\bigl(2(1-q^2),-2(q^2+\eta),-2\eta(1+\eta),\\
 \hspace{6em}-2(1+\eta)^2,0,0\bigr).
\end{gathered}
\end{equation}
Inserting the spinors into the four linear factors in $X$ and $Y$ for each matrix row and clearing denominators reduces these equations to ten polynomial identities in $\mathbb Q[\eta,q]$. Every residual polynomial vanishes coefficientwise. On an open kinematic region where both component matrices have rank five and $v_{0,1}\ne0$, the normalized rows are $w=\wP=v_0/v_{0,1}$. Substitution of Eq.~\eqref{eq:chartdefinitions} gives the invariant form~\eqref{eq:rowlimit}.

At $\eta=2$, $q=4$, the unscaled rational chart gives
$(x,y,z)=(1,2,-1/3)$. The determinants of the submatrices formed from rows 2 through 6 in the numerator convention of Eq.~\eqref{eq:Cdef} are
\begin{equation}\label{eq:minors}
\begin{aligned}
 \det C_{2:6}&=185093021028188160,\\
 \det \CP_{2:6}&=\frac5{20334926626632}.
\end{aligned}
\end{equation}
The factors $[34]$, $\langle56\rangle$, $\langle34\rangle$, and $[56]$ are, respectively, $1$, $-54$, $1$, and $-1/18$. Thus both bases and both normalized row equations are regular. Their common row is
\begin{equation}\label{eq:rowvalue}
 w=\wP=(1,6/5,2/5,3/5,0,0).
\end{equation}
Since the determinants are rational functions and nonzero at this point, rank five holds on a nonempty open part of the chart.

Multiplying all dotted spinors by $r$ rescales every invariant by $r^2$ while preserving masslessness and conservation. Taking $\rho=r^2$ gives the map in Eq.~\eqref{eq:dominance}. Its nonzero Jacobian establishes three independent invariant directions.

\subsection{Momentum-conserving deformation and differentiated row equations}
For the first four legs, deform both spinors linearly using Table~\ref{tab:direction}. Keep $\lambda_5$ and $\lambda_6$ fixed. Define
\begin{equation}\label{eq:rhsconservation}
 Q^{a\dot a}(\tau)=-\sum_{i=1}^4
 \lambda_i^a(\tau)\widetilde\lambda_i^{\dot a}(\tau).
\end{equation}
Momentum conservation at every $\tau$ then fixes
\begin{equation}\label{eq:solveconservation}
\begin{aligned}
 \widetilde\lambda_5^{\dot a}(\tau)
 &=\frac{Q^{0\dot a}(\tau)\lambda_6^1
          -Q^{1\dot a}(\tau)\lambda_6^0}
         {\langle56\rangle},\\
 \widetilde\lambda_6^{\dot a}(\tau)
 &=\frac{\lambda_5^0Q^{1\dot a}(\tau)
          -\lambda_5^1Q^{0\dot a}(\tau)}
         {\langle56\rangle}.
\end{aligned}
\end{equation}
The denominator is fixed and nonzero. These formulas define a momentum-conserving deformation whose dotted spinors are quadratic polynomials in $\tau$ divided by that constant determinant. Each momentum remains massless by its spinor representation.

\begin{table}[tb]
\caption{Spinor variations for legs 1 through 4 defining the momentum-conserving deformation in Appendix~\ref{app:kinematics}. The undotted spinors of legs 5 and 6 are fixed; their dotted spinors are determined by Eq.~\eqref{eq:solveconservation}.}
\label{tab:direction}
\renewcommand{\arraystretch}{1.2}
\centering
\begin{tabular}{ccc}
\hline
Leg & $\delta\lambda_i$ & $\delta\widetilde\lambda_i$\\
\hline
1 & $(0,1)$ & $(1,0)$ \\
2 & $(2,0)$ & $(0,-1)$ \\
3 & $(-1,1)$ & $(1,2)$ \\
4 & $(1,-2)$ & $(-2,1)$ \\
\hline
\end{tabular}

\end{table}

Write $C=C_0+\tau C_1+O(\tau^2)$ and $w=w^{(0)}+\tau w^{(1)}+O(\tau^2)$. The two linear systems to solve are
\begin{equation}\label{eq:rowderivatives}
\begin{aligned}
 w^{(0)}C_0&=0,&(w^{(0)})_1&=1,\\
 w^{(1)}C_0+w^{(0)}C_1&=0,&(w^{(1)})_1&=0,
\end{aligned}
\end{equation}
and the analogous equations for $\CP$. Let $B_0$ and $B_1$ denote rows 2 through 6 of $C_0$ and $C_1$. If $\widehat w$ is the vector formed by entries 2 through 6, then
\begin{equation}\label{eq:explicitrowsolve}
\begin{aligned}
 \widehat w^{(0)}&=-(C_0)_{1,:}B_0^{-1},\\
 \widehat w^{(1)}&=-\left[(C_1)_{1,:}+\widehat w^{(0)}B_1\right]B_0^{-1}.
\end{aligned}
\end{equation}
The determinants in Eq.~\eqref{eq:minors} justify both inverses.

At the stated point the resulting channel derivatives and row difference are
\begin{equation}\label{eq:derivativedata}
\begin{gathered}
 \alpha=-\frac{17}{2},\qquad\beta=\frac{571}{54},\\
 \nu=w^{(1)}-w^{\mathsf P,(1)},\\
 \nu=\left(0,-\frac{1564}{225},-\frac{623}{225},
 -\frac{323}{150},0,0\right).
\end{gathered}
\end{equation}
Here $w^{\mathsf P,(1)}$ is the linear coefficient of the exchanged row. To display the nonzero contraction, Table~\ref{tab:poles} gives
\begin{equation}\label{eq:explicitcontraction}
\begin{aligned}
 K_{\mathcal A}
 &=\left(-\frac{1564}{225}\right)\left(-\frac3{17}\right)
   +\left(-\frac{323}{150}\right)\frac{16}{17}
 =-\frac45,\\
 K_{\mathcal B}
 &=\left(-\frac{623}{225}\right)\frac{45}{571}
   +\left(-\frac{323}{150}\right)\left(-\frac{270}{571}\right)
 =\frac45.
\end{aligned}
\end{equation}
The complete matrix derivatives follow from differentiating Eqs.~\eqref{eq:mdef}-\eqref{eq:Cdef}. The rational point evaluates kinematic coefficients in homogeneous identities, not the resummed four-point amplitude at a massive pole.

\subsection{Cyclic permutations of the factorization channels}
For the two source terms in each component, respectively, the cyclic three-particle channel sets are
\begin{equation}\label{eq:channelsets}
\begin{aligned}
 &(234,123,612,561,456,345),\\
 &(345,234,123,612,561,456).
\end{aligned}
\end{equation}
Momentum conservation equates complementary channels:
$s_{156}=s_{234}$, $s_{126}=s_{345}$, and $s_{456}=s_{123}$.
In the first source term the singular components are therefore
$j=1,3,4,6$, whereas in the second they are $j=1,2,4,5$.
The other four terms have denominator $2y$.

For example, the first term of $Z_1/g^4$ has coefficient
$-(s_{24}/s_{34})/\alpha=(x+y)/(x\alpha)$ and product
$M(y,x)M(z,x+y)=\mathcal A$. The second has coefficient
$-(s_{35}/s_{34})/\beta=(x+y+z)/(x\beta)$ and product
$M(x,y+z)M(y,z)=\mathcal B$. The remaining terms follow from the same substitution using Eq.~\eqref{eq:nonadjacent} and the cyclic rule. This accounts for every entry of Table~\ref{tab:poles}, including its zeros. The restriction to adjacent remainder poles is necessary here: a nonadjacent $s_{13}$ pole would not be regular on the chosen locus.

The common-row and cyclic-pole calculations are identities in independent parameters; Eq.~\eqref{eq:explicitcontraction} establishes nonvanishing of a rational coefficient. The symbolic implementation reconstructs this finite algebra from the definitions above. The extension to every derivative order follows from the argument using the nonzero Jacobian and polynomial identities in Sec.~\ref{subsec:allorders}.

\section{Analytic continuation of the residue expansion}\label{app:analytic}
\subsection{Identification and decay of the coefficients}
Fix $n$ and set $z=1+2u/\mu_n$. Write
$\mathcal F_n(z)=R_n(\mu_n(z-1)/2)$. The endpoint sum
$\sum_\ell b_{n\ell}=r_n<\infty$ and the bound
$|P_\ell(z)|\le1$ on $[-1,1]$ imply uniform convergence of the physical series. Termwise integration and Legendre orthogonality identify
\begin{equation}\label{eq:legendrecoefficients}
 b_{n\ell}=\frac{2\ell+1}{2}
 \int_{-1}^1\mathcal F_n(z)P_\ell(z)\,\dd z.
\end{equation}
Thus the nonnegative coefficients in the assumed representation coincide with the analytic function's orthogonal-expansion coefficients.

The allowed polar set ensures that $\mathcal F_n$ is holomorphic in a neighborhood of $[-1,1]$. Choose a Bernstein ellipse $E_{\rho_0}$, with foci $\pm1$ and parameter $\rho_0>1$, whose closure and a neighborhood lie within that domain. The parametrization is $z=(w+w^{-1})/2$ with $|w|=\rho_0$. Cauchy's formula for the associated Laurent series gives a bound $2B\rho_0^{-j}$ on the degree-$j$ Chebyshev coefficients, where $B$ bounds $|\mathcal F_n|$ on the ellipse. For $\ell\ge1$, the Chebyshev truncation $p_{\ell-1}$ of degree $\ell-1$ therefore satisfies
\begin{equation}\label{eq:approximation}
 \|\mathcal F_n-p_{\ell-1}\|_{[-1,1]}
 \le C\rho_0^{-\ell},
\end{equation}
with $C$ independent of $\ell$.

Orthogonality removes $p_{\ell-1}$ from the integral in Eq.~\eqref{eq:legendrecoefficients}, giving
\begin{equation}\label{eq:decay}
 |b_{n\ell}|\le C(2\ell+1)\rho_0^{-\ell}.
\end{equation}
The $\ell=0$ case is included by increasing $C$.

\subsection{Normal convergence and nonnegative Taylor coefficients}
Choose $1<\rho'<\rho''<\rho_0$. The generating function
\begin{equation}\label{eq:legendregenerating}
 \sum_{\ell=0}^{\infty}P_\ell(z)t^\ell
 =(1-2zt+t^2)^{-1/2}
\end{equation}
is holomorphic for $z$ in the closed smaller ellipse $E_{\rho'}$ and $|t|\le1/\rho''$. Its singularities occur at the roots corresponding to $w$ and $w^{-1}$, outside this $t$ disk. Cauchy's coefficient estimate gives the uniform bound
$|P_\ell(z)|\le B'(\rho'')^\ell$. Combined with Eq.~\eqref{eq:decay}, this majorizes the series by
\begin{equation}\label{eq:majorant}
 CB'\sum_{\ell=0}^{\infty}(2\ell+1)
 \left(\frac{\rho''}{\rho_0}\right)^\ell<\infty.
\end{equation}
The Legendre series is consequently normally convergent on smaller ellipses and represents $\mathcal F_n$ there by the analytic identity theorem. In particular, it can be differentiated term by term in a complex neighborhood of $z=1$.

Returning to $u$, Eq.~\eqref{eq:legendre} gives
\begin{equation}\label{eq:positivetaylor}
 [u^k]R_n(u)=\frac1{\mu_n^k}
 \sum_{\ell\ge k}b_{n\ell}
 \frac{(\ell+k)!}{(\ell-k)!(k!)^2}\ge0.
\end{equation}
The sums converge for fixed $k$ by local normal convergence and its derivative bounds. The Taylor expansion itself converges on $|u|<\mu_1$, because the residue has no allowed singularity in this disk. Its constant term is $r_n>0$, so it is strictly positive for $0<u<\mu_1$. The argument proves exactly the continuation used in Sec.~\ref{sec:spectrum} without imposing finite spin or any positivity statement beyond the first massive transfer pole.

\section{Alternative four-point functions}\label{app:alternatives}
The first two examples concern the reduced four-point reconstruction problem; no full six-point completion is asserted, and they do not establish minimality of the complete assumptions. The last evaluates the obstruction in Sec.~\ref{sec:discussion}.

\subsection{The functional equation without spectral saturation}
Set $g^2=\mu_1=1$. The single-pole example
\begin{equation}\label{eq:onepole}
\begin{aligned}
 M(s,u)&=\frac{1-s-u}{(1-s)(1-u)},\\
 F(s,u)&=-\frac1{su}+\frac1{(1-s)(1-u)}
\end{aligned}
\end{equation}
satisfies Eq.~\eqref{eq:functional} and has
$R_1(u)=1/(1-u)$ and $f(s,0)=1/(1-s)$. It therefore obeys the exact forward identity. Its residue also satisfies partial-wave positivity. In terms of $z=1+2u$, the residue is $2/(3-z)$. Rodrigues' formula and integration by parts imply, for $a>1$,
\begin{equation}\label{eq:residuepositivitycontrol}
 \int_{-1}^1\frac{P_\ell(z)}{a-z}\,\dd z
 =\frac1{2^\ell}\int_{-1}^1
 \frac{(1-z^2)^\ell}{(a-z)^{\ell+1}}\,\dd z>0.
\end{equation}
Taking $a=3$ gives positive Legendre coefficients. Analyticity around the interval ensures that the expansion represents the residue.

Here $a_{00}=a_{20}=1$, so $a_{20}-(2/5)a_{00}^2=3/5$. The function satisfies the reduced functional and spectral conditions but violates Eq.~\eqref{eq:loworder}. Thus those conditions alone do not uniquely determine the Veneziano amplitude.

\subsection{A forward subtraction}
The absence of a subtraction in Eq.~\eqref{eq:dispersion} has a distinct role. At unit scales take
\begin{equation}\label{eq:subtraction}
\begin{gathered}
 h(s)=\frac{e^{-s+cs^2}}{1-s},\qquad
 c=\frac{\sqrt{10}-2}{4},\\
 M(s,u)=\frac{h(s)h(u)}{h(s+u)},\qquad
 F=-\frac{M}{su}.
\end{gathered}
\end{equation}
The function has the stated polar lines, satisfies the functional equation, and obeys
\begin{equation}\label{eq:subtractiondata}
 f(s,0)=\frac1{1-s}+2c,\qquad
 R_1(u)=\frac{e^{-2cu}}{1-u}.
\end{equation}
Since $1+2c=\sqrt{10}/2$, its $a_{00}=1+2c$ and $a_{20}=1$ satisfy the low-order equality.

Its partial-wave coefficients are nonnegative. In terms of $z=1+2u$,
$R_1=2e^c e^{-cz}/(3-z)$. The coefficient of $z^k$ is
\begin{equation}\label{eq:subtractionpowers}
 \frac{2e^c}{3^{k+1}}
 \sum_{j=0}^k\frac{(-3c)^j}{j!}>0,
\end{equation}
because $0<3c<1$ and the alternating terms decrease in magnitude. Each monomial $z^k$ has nonnegative Legendre coefficients: Rodrigues' formula gives a nonnegative integral when $k-\ell$ is even and $k\ge\ell$, and zero otherwise. Uniform convergence of the power expansion on $[-1,1]$ justifies its termwise projection.

The constant $2c$ in Eq.~\eqref{eq:subtractiondata} violates the exact forward identity while preserving the functional equation and low-order equality. Allowing such a subtraction therefore changes the reduced spectral reconstruction.

\subsection{Violation of the functional equation by the entire deformation}
For the deformation in Eq.~\eqref{eq:sinedeformation}, use dimensionless variables with $\mu_1=1$. The first correction to $M$ is $\epsilon\pi^{6N}Q_N(s,u)$, where
\begin{equation}\label{eq:Q}
 Q_N(s,u)=[su(s+u)]^{2N}.
\end{equation}
At $(x,y,z)=(1,2,-1/3)$, the leading difference between the two sides of Eq.~\eqref{eq:functional}, divided by $\epsilon\pi^{6N}$, is
\begin{equation}\label{eq:deformationdefect}
 C_N=6^{2N}+\left(\frac83\right)^{2N}
 -\left(\frac{10}{9}\right)^{2N}
 -\left(\frac{40}{9}\right)^{2N}>0.
\end{equation}
Since $6>40/9>0$ and $8/3>10/9>0$, one has $C_N>0$ for every $N\ge1$. Equation~\eqref{eq:nonzero} gives the nonzero Ward contribution $-(4/5)g^4\epsilon\pi^{6N}C_N$, which no regular term can change by Eq.~\eqref{eq:laurent}.

\bibliographystyle{JHEP}
\bibliography{references}

@article{EHM,
  author = {Elvang, Henriette and Herderschee, Aidan and Morales, Roger},
  title = {{String theory from maximal supersymmetry}},
  journal = {JHEP},
  volume = {07},
  year = {2026},
  pages = {105},
  doi = {10.1007/JHEP07(2026)105},
  eprint = {2601.11705},
  archivePrefix = {arXiv},
  primaryClass = {hep-th},
}

@article{Veneziano,
  author = {Veneziano, Gabriele},
  title = {{Construction of a crossing-symmetric, Regge-behaved amplitude for linearly rising trajectories}},
  journal = {Nuovo Cim. A},
  volume = {57},
  year = {1968},
  pages = {190-197},
  doi = {10.1007/BF02824451},
}

@article{WanZhou,
  author = {Wan, Shi-Lin and Zhou, Shuang-Yong},
  title = {{Analytic Bootstrap of the Veneziano Amplitude}},
  year = {2026},
  eprint = {2605.11084},
  archivePrefix = {arXiv},
  primaryClass = {hep-th},
}

@article{ShaoVichi,
  author = {Shao, Long-Qi and Vichi, Alessandro},
  title = {{Analytic Boundaries of Infinite-Spin-Tower Amplitudes from Hidden Zero}},
  year = {2026},
  eprint = {2607.27300},
  archivePrefix = {arXiv},
  primaryClass = {hep-th},
}

@article{StringsAlmostNothing,
  author = {Cheung, Clifford and Remmen, Grant N. and Sciotti, Francesco and Tarquini, Michele},
  title = {{Strings from Almost Nothing}},
  journal = {Phys. Rev. Lett.},
  volume = {136},
  year = {2026},
  pages = {251601},
  doi = {10.1103/cw4p-cqh7},
  eprint = {2508.09246},
  archivePrefix = {arXiv},
  primaryClass = {hep-th},
}

@article{EFK,
  author = {Elvang, Henriette and Freedman, Daniel Z. and Kiermaier, Michael},
  title = {{Solution to the Ward identities for superamplitudes}},
  journal = {JHEP},
  volume = {10},
  year = {2010},
  pages = {103},
  doi = {10.1007/JHEP10(2010)103},
  eprint = {0911.3169},
  archivePrefix = {arXiv},
  primaryClass = {hep-th},
}

@article{Mansfield,
  author = {Mansfield, Gareth},
  title = {{Positivity of the Veneziano amplitude in ten dimensions}},
  journal = {JHEP},
  volume = {07},
  year = {2025},
  pages = {179},
  doi = {10.1007/JHEP07(2025)179},
  eprint = {2502.20372},
  archivePrefix = {arXiv},
  primaryClass = {hep-th},
}

@article{Carving,
  author = {Huang, Yu-tin and Liu, Jin-Yu and Rodina, Laurentiu and Wang, Yihong},
  title = {{Carving out the space of open-string S-matrix}},
  journal = {JHEP},
  volume = {04},
  year = {2021},
  pages = {195},
  doi = {10.1007/JHEP04(2021)195},
  eprint = {2008.02293},
  archivePrefix = {arXiv},
  primaryClass = {hep-th},
}

@article{Flattening,
  author = {Berman, Justin and Elvang, Henriette and Herderschee, Aidan},
  title = {{Flattening of the EFT-hedron: supersymmetric positivity bounds and the search for string theory}},
  journal = {JHEP},
  volume = {03},
  year = {2024},
  pages = {021},
  doi = {10.1007/JHEP03(2024)021},
  eprint = {2310.10729},
  archivePrefix = {arXiv},
  primaryClass = {hep-th},
}

@article{Corners,
  author = {Berman, Justin and Elvang, Henriette},
  title = {{Corners and islands in the S-matrix bootstrap of the open superstring}},
  journal = {JHEP},
  volume = {09},
  year = {2024},
  pages = {076},
  doi = {10.1007/JHEP09(2024)076},
  eprint = {2406.03543},
  archivePrefix = {arXiv},
  primaryClass = {hep-th},
}

@article{MultiparticleRigidity,
  author = {Arkani-Hamed, Nima and Cheung, Clifford and Figueiredo, Carolina and Remmen, Grant N.},
  title = {{Multiparticle Factorization and the Rigidity of String Theory}},
  journal = {Phys. Rev. Lett.},
  volume = {132},
  year = {2024},
  pages = {091601},
  doi = {10.1103/PhysRevLett.132.091601},
  eprint = {2312.07652},
  archivePrefix = {arXiv},
  primaryClass = {hep-th},
}

@article{HigherSpinConstraints,
  author = {Basile, Ivano and Remmen, Grant N. and Staudt, Georgina},
  title = {{Higher-spin and higher-point constraints on stringy amplitudes}},
  journal = {Phys. Rev. D},
  volume = {114},
  year = {2026},
  pages = {026001},
  doi = {10.1103/nstw-p1p5},
  eprint = {2603.04485},
  archivePrefix = {arXiv},
  primaryClass = {hep-th},
}

@article{HiddenZeros,
  author = {Arkani-Hamed, Nima and Cao, Qu and Dong, Jin and Figueiredo, Carolina and He, Song},
  title = {{Hidden zeros for particle/string amplitudes and the unity of colored scalars, pions and gluons}},
  journal = {JHEP},
  volume = {10},
  year = {2024},
  pages = {231},
  doi = {10.1007/JHEP10(2024)231},
  eprint = {2312.16282},
  archivePrefix = {arXiv},
  primaryClass = {hep-th},
}

@article{AllOrderSplits,
  author = {Arkani-Hamed, Nima and Figueiredo, Carolina},
  title = {{All-order splits and multi-soft limits for particle and string amplitudes}},
  journal = {JHEP},
  volume = {10},
  year = {2025},
  pages = {077},
  doi = {10.1007/JHEP10(2025)077},
  eprint = {2405.09608},
  archivePrefix = {arXiv},
  primaryClass = {hep-th},
}

@article{SplittingIslands,
  author = {Berman, Justin and Elvang, Henriette and Figueiredo, Carolina},
  title = {{Splitting regions and shrinking islands from higher point constraints}},
  journal = {JHEP},
  volume = {10},
  year = {2025},
  pages = {226},
  doi = {10.1007/JHEP10(2025)226},
  eprint = {2506.22538},
  archivePrefix = {arXiv},
  primaryClass = {hep-th},
}

@article{GravitationalParity,
  author = {Berman, Justin and Caron-Huot, Simon and Chandra, Aditi V. and Elvang, Henriette and Herderschee, Aidan and Lin, Loki L. and Morales, Roger},
  title = {{Gravitational Effective Theories with Maximal Supersymmetry and a Peculiar Parity}},
  year = {2026},
  eprint = {2607.14230},
  archivePrefix = {arXiv},
  primaryClass = {hep-th},
}

@article{PositiveMoments,
  author = {Bellazzini, Brando and Elias Mir{\'o}, Joan and Rattazzi, Riccardo and Riembau, Marc and Riva, Francesco},
  title = {{Positive moments for scattering amplitudes}},
  journal = {Phys. Rev. D},
  volume = {104},
  year = {2021},
  pages = {036006},
  doi = {10.1103/PhysRevD.104.036006},
  eprint = {2011.00037},
  archivePrefix = {arXiv},
  primaryClass = {hep-th},
}

@article{TreeUnitarity,
  author = {Arkani-Hamed, Nima and Eberhardt, Lorenz and Huang, Yu-tin and Mizera, Sebastian},
  title = {{On unitarity of tree-level string amplitudes}},
  journal = {JHEP},
  volume = {02},
  year = {2022},
  pages = {197},
  doi = {10.1007/JHEP02(2022)197},
  eprint = {2201.11575},
  archivePrefix = {arXiv},
  primaryClass = {hep-th},
}

@article{StringEFT,
  author = {Chiang, Li-Yuan and Huang, Yu-tin and Weng, He-Chen},
  title = {{Bootstrapping string theory EFT}},
  journal = {JHEP},
  volume = {05},
  year = {2024},
  pages = {289},
  doi = {10.1007/JHEP05(2024)289},
  eprint = {2310.10710},
  archivePrefix = {arXiv},
  primaryClass = {hep-th},
}

@article{EmergentMonodromy,
  author = {Chen, Alan Shih-Kuan and Elvang, Henriette and Herderschee, Aidan},
  title = {{Emergence of String Monodromy in Effective Field Theory}},
  journal = {Phys. Rev. Lett.},
  volume = {133},
  year = {2024},
  pages = {091601},
  doi = {10.1103/PhysRevLett.133.091601},
  eprint = {2212.13998},
  archivePrefix = {arXiv},
  primaryClass = {hep-th},
}

@article{AsymptoticUniqueness,
  author = {Caron-Huot, Simon and Komargodski, Zohar and Sever, Amit and Zhiboedov, Alexander},
  title = {{Strings from massive higher spins: the asymptotic uniqueness of the Veneziano amplitude}},
  journal = {JHEP},
  volume = {10},
  year = {2017},
  pages = {026},
  doi = {10.1007/JHEP10(2017)026},
  eprint = {1607.04253},
  archivePrefix = {arXiv},
  primaryClass = {hep-th},
}
\end{document}